\documentclass[multphys,vecphys]{svmult}

\usepackage{makeidx}         % allows index generation
\usepackage{graphicx}        % standard LaTeX graphics tool
\usepackage{multicol}        % used for the two-column index
\usepackage[bottom]{footmisc}% places footnotes at page bottom
\makeindex             % used for the subject index
\begin{document}

\title*{A new Generation of Spectrometer Calibration Techniques based on Optical Frequency Combs}
\titlerunning{Calibration based on Optical Frequency Combs} %for an abbreviated version of
% your contribution title if the original one is too long
\author{Piet O.~Schmidt\inst{1}\and Stefan Kimeswenger\inst{2}\and Hans Ulrich
K\"aufl\inst{3}}
% Use \authorrunning{Short Title} for an abbreviated version of
% your contribution title if the original one is too long
\institute{Institute of Experimental Physics, Technikerstr. 25, A-6020 Innsbruck
\texttt{Piet.Schmidt@uibk.ac.at}
\and Institute of Astro- and Particle Physics, Technikerstr. 25, A-6020 Innsbruck
\texttt{Stefan.Kimeswenger@uibk.ac.at}
\and European Southern Observatory, Karl-Schwarzschild-Str. 2,
D-85748 Garching
\texttt{hukaufl@eso.org}
}
%
% Use the package "url.sty" to avoid
% problems with special characters
% used in your e-mail or web address
%
\maketitle

Typical astronomical spectrographs \cite{skk:appenzeller,skk:huk2004}
have a resolution $\lambda/\Delta\lambda$ ranging between a few
hundred to 200.000. Deconvolution and correlation techniques are
being employed with a significance down to $1/1000\mathrm{th}$ of a
pixel. HeAr and ThAr lamps are usually used for calibration in low
and high resolution spectroscopy, respectively. Unfortunately, the
emitted lines typically cover only a small fraction of the
spectrometer's spectral range. Furthermore, their exact position
depends strongly on environmental conditions. A problem is the strong
intensity variation between different lines\footnote{see:
www.eso.org/instruments/fors/inst/arc\_lines\_MIT/atlas\_GRIS\_1400V+18.jpeg
-- 587.6 nm He line overexposes while nearly no other line is visible
between 400 and 690 nm} (intensity ratios {$>300$). In addition, the
brightness of the lamps is insufficient to illuminate a spectrograph
via an integrating sphere, which in turn is important to calibrate a
long-slit spectrograph, as this is the only way to assure a uniform
illumination of the spectrograph pupil.

Laboratory precision laser spectroscopy has experienced a major
advance with the development of optical frequency combs generated by
pulsed femto-second lasers. These lasers emit a broad spectrum
(several hundred nanometers in the visible and near infra-red) of
equally-spaced "comb" lines with almost uniform intensity (intensity
ratios typically $<10$). Self-referencing of the laser establishes a
precise ruler in frequency space that can be stabilized to the
10$^{-18}$ uncertainty level \cite{skk:stenger,skk:zimmermann},
reaching absolute frequency inaccuracies at the 10$^{-12}$ level per
day when using the Global Positioning System's (GPS) time signal as
the reference. The exploration of the merits of this new technology
holds the promise for broad-band, highly accurate and reproducible
calibration required for reliable operation of current and next
generation astronomic spectrometers. Similar techniques are also
proposed in \cite{skk:constanza,skk:murphy}.

\section{Calibration of high resolution spectrometers} We will
consider optical frequency combs based on fiber lasers that have a
repetition frequency $f_\mathrm{rep}\sim 250$~MHz, therefore
producing an equally spaced spectrum with lines separated by 250 MHz.
These systems have the advantage over Ti:Sapphire based frequency
combs to be more reliable and require less maintenance. To resolve
individual lines of the frequency comb, a resolution of more than
$2\times 10^{-6}$, would be required. Therefore, astronomical
spectrographs will see these devices as white light. For the
calibration of high resolution spectrometers, we propose to filter
the output of a frequency comb generator with external cavities as
shown in Fig. \ref{skk:fig:1}(a). Interference inside the cavity
leads to a frequency dependent transmission. The separation of
transmission maxima (free spectral range: $\Delta f_\mathrm{FSR}$)
can be chosen via the length of the cavity, whereas the width of the
maxima is determined by the reflectivity of the optical coating
applied to the surfaces of the cavity mirrors.
%We can choose these parameters in such a way that a whole set
%of neighboring comb lines will be transmitted e.g. every 550 GHz
%($\equiv\,\,\Delta\lambda$=0.6nm @ 600nm). These devices can be
%easily varied by several orders of magnitude.
\begin{figure}
\centering
\includegraphics[width=\columnwidth]{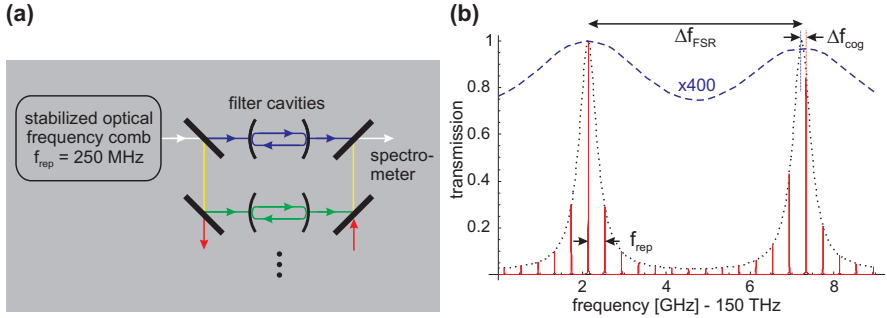}
\caption{(a) Schematic setup of the calibration source: the frequency
comb provides a stabilized ruler in frequency space that is filtered
by an external cavity. The distance between transmission maxima of
the cavity can be adjusted to match the resolution of the
spectrometer. (b) Cavity-filtered output spectrum of a frequency comb
(exaggerated for visibility). Solid line: filtered comb spectrum;
dotted line: transmission curve of the cavity; dashed line: spectrum
seen by the spectrometer (magnified by $400$).}
\label{skk:fig:1}       % Give a unique label
\end{figure}

By matching the repetition frequency of the optical comb to be an
integer multiple of the cavity's free spectral range ($\Delta
f_\mathrm{FSR} = n f_\mathrm{rep}$), the effective repetition
frequency of the laser is increased to $\Delta f_\mathrm{FSR}$. If
the transmission maxima have a spacing well exceeding the
spectrometer's resolution, and the frequency of one of the observed
lines can be identified unambiguously, one can assign a precise
frequency to all other observed lines simply by counting. This
identification can be achieved by overlapping a cw laser (referenced
to the frequency comb via one of the transmitted comb lines) with the
comb spectrum before the filter cavity and observing the light of
this laser on the spectrometer. At the same time, this laser serves
as a reference to actively stabilize the length of the filter cavity
and therefore its spectral properties. Due to limitations in the
optical coatings, it is unlikely that a single filter cavity can
cover the entire spectral range. A solution would be to spectrally
split the output of the frequency comb laser into several wavelength
regions, each filtered by an optimized cavity. The filtered output of
the frequency comb will be similar to the solid line in Fig.
\ref{skk:fig:1}(b). The recorded spectrum is a convolution between
the filtered spectrum of the frequency comb and the spectrometer's
resolution (dashed line in Fig. \ref{skk:fig:1}(b)). The achievable
quality of the optical coating (dispersion compensated bandwidth vs.
reflectivity/absorption) determines the width of the transmission
resonances. This may result in insufficient suppression of
neighboring comb lines and thus shift the center of gravity of the
line observed by the spectrometer. It will be difficult to exactly
match the filter cavity's free spectral range to the repetition
frequency of the optical comb over the whole spectral range due to
residual dispersion effects. This will result in an imperfect match
of the comb lines to the transmission maxima of the filter cavity and
thus induce a shift in the observed line center by $\Delta
f_\mathrm{cog}$ as shown in Fig. \ref{skk:fig:1}(b)\footnote{Such a
shift may not be an issue as long as it is reproducible and absolute
frequency accuracy is not required.}.

In the following, we will give an estimate of the shift in line
center calibration due to uncompensated dispersion in the filter
cavity based on CRIRES (2 pixel Nyquist sampling resolution
$\lambda/\Delta\lambda=10^5\equiv$ 1.5 GHz @ $\lambda=2\mu$m).
Emission line centers can currently be determined to within 0.05
pixels (0.001 pixels $\equiv$ 0.75 MHz anticipated in future
experiments) \cite{skk:huk2007}. To achieve accurate fitting of the
center of gravity of the maxima, a separation of $\approx$ 27 pixels
is required. This corresponds to $\Delta f_\mathrm{FSR} = 20$ GHz.
%We define the shift in the center of gravity of the observed
%cavity transmission line as follows:
%\[
%\Delta f_n^\mathrm{cog}=\frac{\int\limits_{-\Delta
%f_\mathrm{FSR}/2}^{\Delta f_\mathrm{FSR}/2} I_\mathrm{cav}(f-f_n)
%I_\mathrm{comb}(f-f_n)(f)\,df}{\int\limits_{-\Delta
%f_\mathrm{FSR}/2}^{\Delta f_\mathrm{FSR}/2} I_\mathrm{cav}(f-f_n)
%I_\mathrm{comb}(f-f_n)\,df}
%\]
Fig. \ref{skk:fig:2} shows the effect of a frequency shift between a
comb line and a filter cavity resonance. The intensity of a single
transmission maximum as seen by the spectrometer is periodic in
$f_\mathrm{rep}$ (several comb lines contribute to each maximum with
decreasing intensity as their distance from the maximum increases).
The observed shift in the center of gravity exhibits plateaus as comb
lines approach the transmission maximum of the cavity. From the inset
in Fig. \ref{skk:fig:2}(b) we see that a maximum frequency shift
between comb line and cavity resonance of 18 MHz can be tolerated to
maintain the required line center accuracy of 0.75 MHz. Fig.
\ref{skk:fig:2}(a) shows that at this shift, the intensity has
reached 7\% of its maximum value. This defines a threshold intensity
below which the observed line is discarded for calibration purposes.
Since the spectrum of the frequency comb can not be assumed uniform
in intensity to that level, a scan of the frequency shift between
cavity resonances and frequency comb teeth is required to calibrate
the magnitude of the transmission maxima for each cavity resonance
\cite{skk:schliesser}.

\begin{figure}
\centering
\includegraphics[width=300pt]{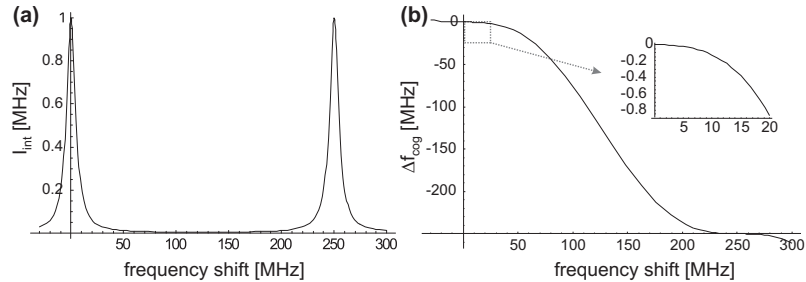}
\caption{Effect of frequency shift between filter cavity resonance
and frequency comb lines. (a): Intensity as seen by the spectrometer,
integrated over one cavity transmission maximum. (b): Shift in the
center of gravity of the observed line (simulation parameters:
$f_\mathrm{rep} = 250$ MHz, $\Delta f_\mathrm{FSR} = 20$ GHz, Finesse
$= 2000$).}
\label{skk:fig:2}       % Give a unique label
\end{figure}

Even more stringent requirements in terms of reproducibility and
resolution apply to e.g. HARPS and CODEX: At similar resolutions the
required stability over a few months to several years has to be
$10^{-3}\dots 10^{-5}$ pixels \cite{skk:lovis_pepe}.

\section{Calibration of mid resolution spectrometers} For medium
resolution spectrometers with an effective resolution below 100.000,
the stability of the frequency comb is not required. Instead, it can
be replaced by a fiber laser-based high-brightness white light source
that is then filtered by the cavities. In this case, the cavity
transmission maxima provide the ruler required for spectrometer
calibration. The dispersion properties of the cavity can be
calibrated using a frequency comb as described in
\cite{skk:schliesser}. We propose to use a cw laser locked to a
stable reference (gas cells or a GPS-referenced frequency comb) to
stabilize the length of the filter cavity to sub-MHz precision. A
similar technique has been successfully implemented previously using
an unstabilized cavity \cite{skk:bacon,skk:foltz}.

%%%%%%%%%%%%%%%%%%%%%%%%%%%%%%%%%%%%%%%%%%%%%%%%%%%%%%%%%%%%%%%%%%%%%%  }

%%%%%%%%%%%%%%%%%%%%%%%%%%%%%%%%%%%%%%%%%%%%%%%%%%%%%%%%%%%%%%%%%%%%%%

\printindex

\begin{thebibliography}{99.}
\bibitem{skk:appenzeller} I. Appenzeller, K. Fricke, W. Furtig, et al:
The Messenger \textbf{94}, pp. 1--6 (1998)
\bibitem{skk:huk2004} H.U. K\"aufl, P. Ballester, P. Biereichel, et al: Ground-based
Instrumentation for Astronomy. In: \textit{Proceedings of the SPIE},
vol 5492, ed. by A.F.M. Moorwood and M. Iye (SPIE, Glasgow, Scotland,
United Kingdom 2004) pp. 1218--1227
\bibitem{skk:stenger} J. Stenger, H. Schnatz, C. Tamm, H.R. Telle: Phys. Rev. Lett. \textbf{88}, 073601-1-4 (2002)
\bibitem{skk:zimmermann} M. Zimmermann, C. Gohle, R. Holzwarth, et al: Opt. Lett. \textbf{29}, 310 (2003)
\bibitem{skk:constanza} C. Araujo-Hauck, L. Pasquini, A. Manescau, et
al: this volume (2007)
\bibitem{skk:murphy} M. T. Murphy, Th. Udem, R. Holzwarth, et al: arXiv:astro-ph/0703622
(2007)
\bibitem{skk:huk2007} H.U. K\"aufl, et al: this volume (2007)
\bibitem{skk:schliesser} A. Schliesser, C. Gohle, T. Udem, T.W. H\"ansch: Optics Express \textbf{14}, 5975 (2006)
\bibitem{skk:lovis_pepe} C. Lovis, F. Pepe: this volume (2007)
\bibitem{skk:bacon} R. Bacon, Y. Georgelin, G. Monnet: Bull. CFHT, \textbf{23} (1990)
\bibitem{skk:foltz} C.B. Foltz, F.H. Chaffee, D.B. Quellette, et al: MMT Tech. Mem. \textbf{85-4} (1985)
\end{thebibliography}
\end{document}